\documentclass[%
 reprint,
superscriptaddress,
showpacs,preprintnumbers,
 amsmath,amssymb,
aps,
prl,
]{revtex4-1}

\usepackage{comment}

\usepackage{txfonts}
\usepackage{amsfonts}
\usepackage{color}%
\usepackage{graphicx}% Include figure files
\usepackage{dcolumn}% Align table columns on decimal point
\usepackage{bm}% bold math
\begin{document}

%\preprint{APS/123-QED}

\title{
Weyl Node and Spin Texture in Trigonal Tellurium and Selenium 
}% Force line breaks with \\

\author{Motoaki Hirayama}
%\email{m-hirayama@aist.go.jp}
\affiliation{%
 Nanosystem Research Institute, AIST, Tsukuba 305-8568, Japan
}%
\affiliation{%
 Department of Physics, Tokyo Institute of Technology, Ookayama, Meguro-ku, Tokyo 152-8551, Japan 
}%
\affiliation{%
TIES, Tokyo Institute of Technology, Ookayama, Meguro-ku, Tokyo 152-8551, Japan 
}%

\author{Ryo Okugawa}
\affiliation{%
 Department of Physics, Tokyo Institute of Technology, Ookayama, Meguro-ku, Tokyo 152-8551, Japan 
}%

\author{Shoji Ishibashi}
\affiliation{%
 Nanosystem Research Institute, AIST, Tsukuba 305-8568, Japan
}%

\author{Shuichi Murakami}
\affiliation{%
 Department of Physics, Tokyo Institute of Technology, Ookayama, Meguro-ku, Tokyo 152-8551, Japan 
}%
\affiliation{%
TIES, Tokyo Institute of Technology, Ookayama, Meguro-ku, Tokyo 152-8551, Japan 
}%

\author{Takashi Miyake}
\affiliation{%
 Nanosystem Research Institute, AIST, Tsukuba 305-8568, Japan 
}%

\date{September 26, 2014}% It is always \today, today,

\begin{abstract}
We study Weyl nodes in materials with broken inversion symmetry.
We find based on first-principles calculations that trigonal Te and Se have multiple Weyl nodes near the Fermi level. 
The conduction bands have a spin splitting similar to the Rashba
splitting around the H points, but unlike the Rashba splitting the spin directions are radial, forming a hedgehog spin texture around the H points, with a nonzero Pontryagin index for each spin-split conduction band.
The Weyl semimetal phase, which has never been observed in real materials without inversion symmetry, is 
realized under pressure.
The evolution of the spin texture by varying the pressure can be explained by the evolution of the Weyl nodes in $\bm{k}$ space.
\end{abstract}

\pacs{71.20.Mq, 71.70.Ej, 03.65.Vf, 71.30.+h}

\maketitle

%%%%%%%%%%%%%%%%%%%%%%%%%%%%%%%%%%%%%%%%%%%%%%%%%%%%%%%%%%%%%%%%%%%%%%%%%%%%%%%%%%%%%%%%%%

Materials having a linear dispersion (Dirac cone) such as graphene~\cite{neto09} have been under intensive investigation recently.
Among various Dirac cones in band structures of solids, three-dimensional Dirac cones
without spin degeneracy are of particular interest because of their topological nature. 
A material having three-dimensional Dirac dispersion without degeneracy near the 
Fermi energy is called a Weyl semimetal~\cite{Wan}.
In Weyl semimetals the Dirac cones have no spin degeneracy because of the spin-orbit interaction (SOI).
The gapless Dirac points without degeneracy are called Weyl nodes.
Interestingly, the Weyl nodes are protected topologically against small perturbations because they have nontrivial topological invariants, namely a monopole charge associated with the Berry curvature in $\bm{k}$ space~\cite{Berry84,Volovik,Murakami07b}.
Each Weyl node is either a monopole or an antimonopole, having a monopole charge of $+1$ or $-1$, respectively.
The effective Hamiltonian around the Weyl node is expressed by a $2 \times 2$ matrix which has degrees of freedom of helicity reflecting the monopole charge.
Thus the system having Weyl nodes naturally has nontrivial spin texture.
In particular, if the lattice structure lacks mirror symmetry,
the Weyl node will bring unconventional spin texture and nontrivial phenomena, because mirror symmetry imposes strong restrictions on the spin direction.
The Weyl semimetal is realized in a system which breaks time-reversal or inversion symmetry.
For Weyl semimetals with broken time-reversal symmetry, pyrochlore iridates~\cite{Wan,Xu}, HgCr$_2$Se$_4$~\cite{Yang}, 
and a superlattice with a normal insulator and a topological insulator with magnetic doping~\cite{BurkovPRL, BurkovPRB} have been proposed. 
On the other hand, for Weyl semimetals with broken inversion symmetry, a superlattice consisting of a normal insulator and a topological insulator with an external electric field is proposed~\cite{Halasz}, while there have been no proposals of real materials for Weyl semimetals without inversion symmetry.

\begin{figure}[ptb]
\centering 
\includegraphics[clip,width=0.45\textwidth ]{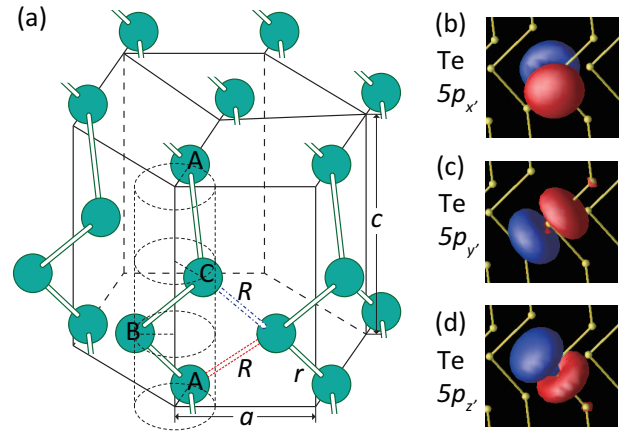} 
\caption{(Color online) (a) Crystal structure of trigonal Te (Se) 
in the $P3_{1}21$ structure.
(b)-(d) Isosurface of the maximally localized Wannier functions %$\pm 0.05$ a.u. 
for the Te-$5p$ orbitals.
}
\label{structure_Wannier}
\end{figure} 

In this letter, we will show using first-principles calculations that 
trigonal Te and Se are such systems.
The materials are gapped at ambient pressure. 
We find spin-split conduction bands near the Fermi energy, where the bands are degenerate at the H point, and the splitting is linear in the wavevector.
This degeneracy has a unit monopole charge for the Berry curvature, and can be regarded as a Weyl node.
The spin around the H point is hedgehog-like radial texture, which reflects low symmetry of the lattice structure.
It is in contrast with the typical spin splitting where spin and velocity are perpendicular to each other, observed in Rashba systems~\cite{Hoesch,ishizaka11} and in topological insulators~\cite{Hsieh}.
We also find that trigonal Te shows the Weyl semimetal phase with time-reversal symmetry under pressure.
The hallmark of the Weyl semimetal phase can be experimentally observed by e.g. the temperature dependence of various physical quantities such as specific heat, compressibility, diamagnetic susceptibility, and dc conductivity~\cite{yang13}.

Te and Se have a characteristic helical structure, as shown in Fig.~\ref{structure_Wannier} (a).
The helical chains containing three atoms in a cell 
are arranged in a hexagonal array. 
The space group is $P3_{1}21$ or $P3_{2}{21}$ ($D^{4}_{3}$ or $D^{6}_{3}$) 
depending on the right-handed or left-handed screw axis.
The materials are $p$ electron systems (Figs.~\ref{structure_Wannier}(b)-(d)).
They are semiconducting at ambient pressure.
A transition to metal takes place under pressure.
The trigonal phase is stable at about 4 and 14 GPa at room temperature in Te and Se, respectively,
and the pressure of the structural phase transition in Te increases by about 0.03-0.04 GPa as the temperature decreases down to 10 K~\cite{blum65}.
As far as we know, there have been no studies on the metal-insulator transition and the structural transition in the trigonal phase at low temperature under high pressure.
The electronic structure has been studied
using the $\bm{k} \cdot \bm{p}$ perturbation method~\cite{doi70} and 
the pseudopotential technique~\cite{joannopoulos75,starkloff78}.
A recent density-functional calculation including the SOI predicted that Te would become a strong topological insulator under shear strain~\cite{agapito13}.

A structural phase transition is experimentally observed near the pressure of metal-insulator transition.
However, the high pressure phase is still under debate~\cite{hejny04,marini12}.
The relation between the metal-insulator transition and structural phase transition is not yet clarified as well.
In this letter, we study the pressure range up to the vicinity of structural phase transition, 3.82 GPa in Te and 14.0 GPa in Se~\cite{se14gpa}, in which the trigonal structure holds even at room temperature.
The trigonal phase is more stable at lower temperature.
Our calculations are based on relativistic density function theory using QMAS code~\cite{qmas},
and the $GW$ correction is added~\cite{schilfgaarde06,miyake08,sm_gwso}.

\begin{figure}[ptb]
\centering 
\includegraphics[clip,width=0.45\textwidth ]{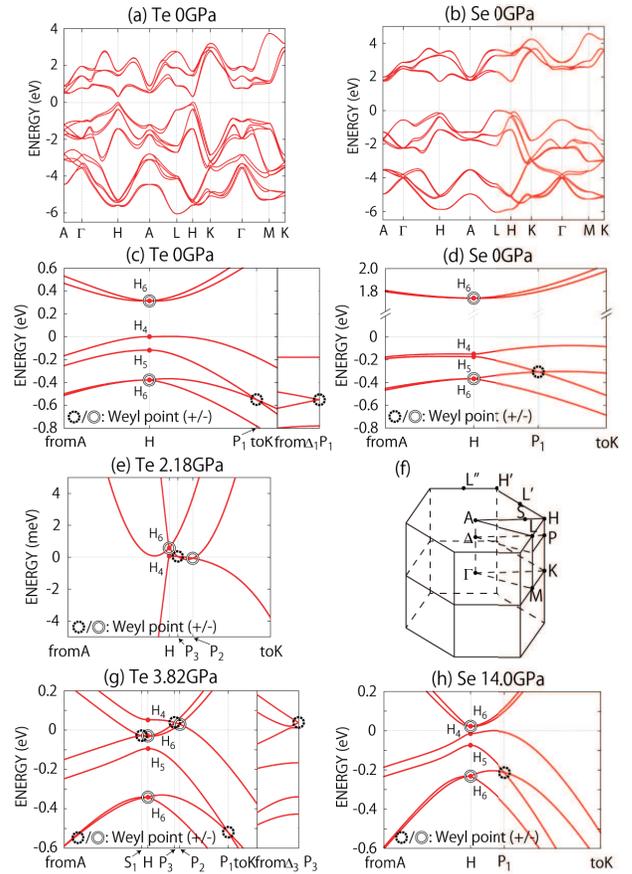} 
\caption{(Color online)
Electronic structures of (a) Te and (b) Se at ambient pressure obtained by the $GW$+SO.
(c) and (d) are magnified figures for Te and Se, respectively.
(e) and (g) are electronic structures of Te at $2.18$ GPa and $3.82$ GPa, respectively, and (h) is that of Se at $14.0$ GPa.
(f) is the Brillouin zone.
P$_1$ (P$_3$) on the H-K line and $\Delta _1$ ($\Delta _3$) on the A-$\Gamma$ line have the same $k_z$.
S$_1$ (P$_2$) is on the H-A (H-K) line.
Weyl points are indicated by the black dashed/white solid circles having a positive/negative monopole charge
(calculated from lower eigen states around the degenerate point).
The energy is measured from the Fermi level.
}
\label{bndTeSe}
\end{figure} 

Figures~\ref{bndTeSe}(a) and (b) show the band structures of Te and Se at ambient pressure.
(See Fig.~\ref{bndTeSe}(f) for the symmetry points.)
Both materials have 18 states near the Fermi level ranging from $-$6 to +4 eV. 
The 18 eigenstates are classified into three types~\cite{joannopoulos75}. 
The deepest six states have bonding character between the nearest-neighbor atoms. 
The middle six states are lone-pair states, 
and the unoccupied six states are anti-bonding states.
Both the bonding and anti-bonding states mainly originate from the $p_{y'}$ and $p_{z'}$ orbitals
($(|p_{y'i} \rangle +|p_{z'(i+1)} \rangle )/\sqrt{2} $ and $(|p_{y'i} \rangle -|p_{z'(i+1)} \rangle )/\sqrt{2} $, respectively),
while the lone-pair bands mainly consist of the $p_{x'}$ orbitals.
Here, $|p_{y'i} \rangle$ ($|p_{z'i} \rangle$), shown in Fig.~\ref{structure_Wannier}(c) (Fig.~\ref{structure_Wannier}(d)), 
is centered at the $i$-th site and oriented to the nearest-neighbor atom in the positive (negative) $c$ direction. 
The remaining orbital, shown in (b), is extended perpendicular to the triangle formed by the two bonds.
We call this orbital $p_{x'}$.
The nearest-neighbor hopping between the $p_{y'}$ and $p_{z'}$ orbitals reaches 
2.4 eV (3.3 eV) in Te (Se). 
This strong $\sigma$ bonding generates a large energy separation between the bonding and anti-bonding bands, 
which contributes to the stabilization of the helical structure.
The materials are gapful, and 12 out of the 18 bands are occupied.
The band gap in the $GW$+SO is $0.314$ $(1.74)$ eV in Te (Se), 
which is in good agreement with the experimental value of $0.323$ $(2.0)$ eV~\cite{anzin77,tutihasi67}.
Both the bottom of the conduction band and top of the valence band 
are close to but slightly off the H point in Te (Fig.~\ref{bndTeSe}(c)). 
In Se, the top of the valence band is located near the L point, and 
the energy level at H is slightly lower than L (Fig.~\ref{bndTeSe}(d)).
The lowest unoccupied state (LUS) at H is doubly-degenerate. 
Their character is $(|p_{y'i} \rangle +|p_{z'i} \rangle )/\sqrt{2} $.
The highest occupied state (HOS) at H has strong $p_{x'}$ character, 
with small $p_{y'}$ and $p_{z'}$ components.

As the Fermi level is shifted across the LUS level, the Fermi surface (FS) shrinks to the H point, then turns to grow larger. 
We calculate the monopole charge of the LUS at H~\cite{sm_mc},
and find that the doubly-degenerate states form a Weyl node.
The same analyses for other degenerate states indicate 
that some of them at H, K, A, $\Gamma$, L and M are also Weyl nodes with monopole charge $\pm 1$.
In the valence bands on the H-K line, there is an accidental degenerate point P$_1$.
This point is also Weyl node.
Of particular interest are the Weyl nodes at the H points close to the 
Fermi energy as the pressure is increased.
The H points are not the time-reversal invariant momenta, but invariant under the operations of both a rotation of $2\pi /3$ about the $c$ axis followed by the fractional translation $\bm{c}/3$, $S_3$, and a rotation of $\pi$ about the $a$ axis, $C_2$.
At the H points the irreducible representations (irreps.) 
are two one-dimensional irreps., H$_4$ and H$_5$,  and a two-dimensional irrep., H$_6$ 
\cite{doi70}.
The state with the H$_6$ irrep.\ is always a Weyl node~\cite{sm_wn},
as one can see from 
the two-band Hamiltonian for the doubly-degenerate occupied bands near the H point obtained in Ref. \cite{doi70}.

\begin{figure}[ptb]
\centering 
\includegraphics[clip,width=0.45\textwidth ]{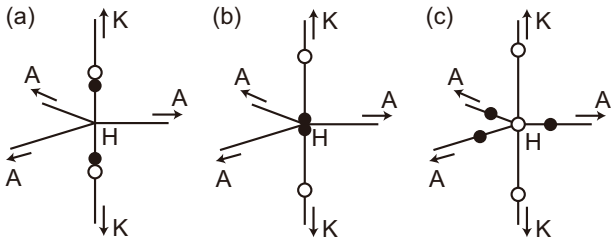} 
\caption{
Motion of Weyl nodes by increasing pressure. The filled and empty circles represent
monopoles and antimonopoles for the HOS, respectively. (a) When pressure is increased, a monopole-antimonopole pair is created at each P point (P$_2$ and P$_3$ in Fig.~\ref{bndTeSe}(e)) at 2.17 GPa, and the system becomes a Weyl semimetal.  
They run along the H-K lines, and (b) two monopoles meet at each of the H points at 2.19 GPa.
(c) Then they evolve into three monopoles along the H-A lines (S$_1$ in Fig.~\ref{bndTeSe}(g)), and one antimonopole residing at the H point.
}
\label{Weyl}
\end{figure}

The band gap decreases with increasing pressure.
The bottom of the conduction band is pulled down toward the Fermi level, 
and the shape of the conduction band bottom becomes sharp as the insulator-to-metal transition is approached.
This is consistent with the experiment that the thermal dependence of the resistivity decreases with increasing pressure~\cite{blum65}.
In Te, when the pressure is increased, the gap closes at four P points, which are related with each 
other by the $C_2$ rotation and time-reversal symmetry.
Each of this gap closing at P corresponds to a monopole-antimonopole pair creation of Weyl nodes,
and the system becomes a Weyl semimetal at 2.17-2.19 GPa (Fig.~\ref{bndTeSe}(e)).
The P$_2$ and P$_3$ points are Weyl nodes between the valence and conduction bands.
Topological surface states connecting these points (Fermi arcs) emerge on the $(01\bar{1}0)$ surface~\cite{sm_tss}.
It is seen by constructing an effective model around these points,
noting the irreducible representations at the P points: the HOS with P$_6$ and the LUS with P$_4$ or P$_5$.
Although the transition pressure could be slightly affected by the numerical convergence,
the existence of the Weyl semimetal phase is robust because the Weyl node is stable for small perturbations in a three-dimensional system with broken inversion symmetry~\cite{Murakami07b}.
At higher pressure, the system becomes metallic.
In addition to P$_1$, the P$_2$ and P$_3$ points on the H-K line and the S$_1$ point on the H-A line are also Weyl nodes (Fig.~\ref{bndTeSe}(g)).

In Se under pressure ($\sim$1 GPa), the highest occupied energy level at H becomes higher than that at L, and the top of the valence band is located near the H point.
The shape of the 18 $p$ bands at 14.0 GPa is very similar to that of Te at 3.82 GPa, but Se is still insulating with an energy gap of 0.02 eV (Fig.~\ref{bndTeSe}(h)) because the SOI in Se is weaker than that in Te.
The splitting between H$_4$ and H$_5$ becomes larger than that at ambient pressure because the splitting originates from the inter hopping between the chiral Se chains.
With higher pressure keeping the trigonal structure, the four Weyl nodes will emerge near the H points under the Fermi level.

Motion of Weyl nodes of Te under the phase transitions are schematically shown in Fig.~\ref{Weyl}(a).
In the entire Brillouin zone, there are eight Weyl points, i.e. four monopoles and four antimonopoles for the HOS.
As we increase the pressure further, the monopoles move and eventually, 
at each H point two monopoles meet  (Fig.~\ref{Weyl}(b)).
This corresponds to the crossing of the doubly degenerate LUS (H$_6$) at H and the single HOS (H$_4$), by increasing pressure.
In fact, there emerge other small FS's in addition to the Weyl nodes on the H-A line,
and furthermore, these Weyl nodes are not on the same energy as seen in Fig.~\ref{bndTeSe}(e).
Nonetheless, the energy bands are gapped between the monopole and antimonopole along 
the H-K line and therefore the topological properties of Weyl semimetals survive.
When the pressure is increased, these two monopoles at each H point are dissociated into three monopoles and one antimonopole,
and the three monopoles run along the three H-A lines  (Fig.~\ref{Weyl}(c)). 
This dissociation necessarily occurs because the HOS at H now belongs to the H$_6$ irrep.\ 
which imposes a single Weyl point at the H point. 
On the other hand, the four antimonopoles at the P points remain, but their energies becomes 
higher than that at H points.
At this pressure, the system is a metal (Fig.~\ref{bndTeSe}(g)), and 
an electron-like pocket appears around the H point.

\begin{figure}[ptb]
\centering 
\includegraphics[clip,width=0.45\textwidth ]{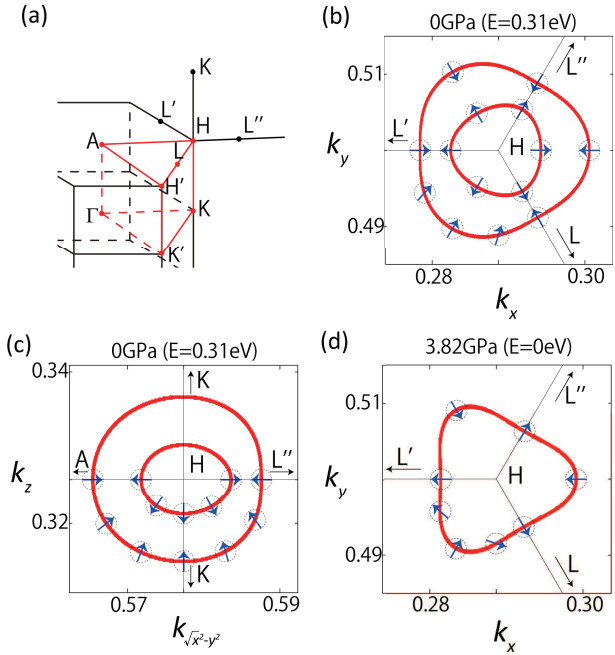} 
\caption[]{(Color online)
Fermi surface and spin structure of Te.
(a) is the Brillouin zone around the H point.
The $\bm{k}$ point on the red line is unchanged under the $S_3$ or $C_2$ operations. 
The Fermi level is shifted by 0.31 eV in (b) and (c). 
(d) is the Fermi surface at 3.82 GPa. 
Arrows surrounded by circles represent the spin projected onto each plane.
}
\label{fsTe}
\end{figure}

Finally, we discuss spin texture in $\bm{k}$ space.
The spin is parallel to the side on the sides of the triangular prism formed by 
the A, H, H', $\Gamma$, K, and K' points (Fig.~\ref{fsTe}(a)), because of the symmetry reason for 
the $S_3$ operation (for the A-$\Gamma$, H-K and H'-K' lines) and 
$C_2$ operation (for other lines).
Figures~\ref{fsTe} (b)(c) are the FS's and the spin textures in Te at ambient pressure.
(Since the system is insulating, we have slightly shifted the Fermi level by 0.31 eV.)
We can see two FS's around the H point.
Remarkably, the spin is oriented radially, and rotates once around H.
The direction on the small FS is opposite to that in the large FS.
In fact, although the energy splitting of the FS's looks similar to the Rashba splitting, the radial structure of spins is different.
In usual Rashba systems in metal surfaces, semiconductor heterostructures 
or three-dimensional Rashba systems such as BiTeI~\cite{ishizaka11},
mirror symmetries are present, and the spins become perpendicular to
the mirror plane when the wavevector is on the mirror plane, and radial spin texture is
never realized.
In addition, the two FS's are separated with each other in Te and Se.
If the present system had mirror symmetry in addition, whose mirror plane includes the $z$ axis, the two FS's would touch each other
at $k_x=k_y=0$ by symmetry.
The completely separated three-dimensional FS's are characteristic of the system without mirror symmetry.
Thus the small (large) FS around the H point has a hedgehog spin structure, characterized by a Pontryagin index $-1$ ($+1$)~\cite{sm_st}.

We also note that in ambient pressure, the radial spin texture for the LUS is identical between the two H points, H and H'.
This spin splitting due to the SOI is represented by a Hamiltonian $\Lambda \bm{\sigma}\cdot\bm{k}$ where $\bm{k}$ is a deviation of the wavevector from H or H', $\bm{\sigma}$ is the spin, and $\Lambda$ is a constant.
This spin splitting leads to spin transport such as spin Hall effect and current-induced spin polarization, whose spin directions are different from the conventional Rashba splitting.
In addition, the HOS at H and that at  H' have opposite spins along the $z$ direction.
This is reminiscent of the spin splitting in MoS$_2$ and WS$_2$ thin films~\cite{xiao12}.
Therefore one can associate the p-type carriers at the H and H' points with a valley degree of freedom, which couples with the spins.

Because the two structures  $P3_121$ and $P3_221$ are transformed into each other by space inversion, they realize opposite signs for the spin-valley coupling.
Hence a domain wall between the two structures may potentially be used as a ``valley switch'' for p-type carriers. On the other hand, the LUS shows opposite spin textures between the two crystallographic structures, and they correspond to the two signs of the SOI term $\pm\Lambda \bm{\sigma}\cdot\bm{k}$.

In the metallic phase at 3.82 GPa, there are three FS's near the H point (see Fig.~\ref{bndTeSe}(g)).
One is surrounding the H point, and is electron-like. The other two are hole-like. 
The former cut at $k_z=\pi/c$ is shown in Fig.~\ref{fsTe}(d).
The spin has no $z$ component on this plane.
It is oriented radially on the H-L, H-L', and H-L'' lines, but this is not the case at other $\bm{k}$ points.
As we follow the FS on the $k_{z}=\pm \pi /c$ plane and go around the H point, the spin rotates twice.
This reflects the fact that within the $k_xk_y$-plane 
the sum of monopole charge amounts to two (Fig.~\ref{Weyl}(c)).
On the other hand, the spin on the hole-like FS's rotates only once on the $k_xk_y$-plane.
In the $P3_221$ structure, the direction of spin is opposite to that in $P3_121$.

In summary, we have studied the electronic structures of trigonal Te and Se as the systems having the Weyl nodes without inversion symmetry.
The materials undergo insulating-semimetallic-metallic transitions under pressure.
We find this semimetallic phase is a Weyl semimetal,
which is a first proposal for real materials of Weyl semimetals with broken inversion symmetry~\cite{cm_liu14}.
The phase transition is fully understood in terms of the Weyl nodes thanks to their topological nature.
The conduction band at ambient pressure shows a Rashba-like spin splitting near the H points, but shows a hedgehog spin texture unlike the Rashba splitting.
The spin is directed to the side on the $\Gamma$KK'-AHH' triangular prism, 
and rotates around the H point.
Under pressure, the number of the rotations varies on the $k_xk_y$-plane.

This work was supported by Grant-in-Aid for Scientific Research
(No. 22104010, No. 24540420 and No, 26287062),
by the Computational Materials Science Initiative (CMSI)
and the Strategic Programs for Innovative Research (SPIRE), Mext, Japan,
and also by MEXT Elements Strategy Initiative to Form Core Research Center (TIES).

%%%%%%%%%%%%%%%%%%%%%%%%%%%%%%%%%%%%%%%%%%%%%%%%%%%%%%%%%%%%%%%%%%%%%%%%%%%%%%

\clearpage
\noindent
{\Large Supplemental Materials}

\renewcommand{\theequation}{S.\arabic{equation}}
\setcounter{equation}{0}
\renewcommand{\tablename}{Table S}
\setcounter{table}{0}
\renewcommand{\figurename}{Fig. S}
\setcounter{figure}{0}

\noindent
\section*{S.1 Detail of \textit{ab initio} Calculation}

Our calculations are based on the density function theory in the local density approximation (LDA)~\cite{S_perdew81}. 
We use a first-principles code QMAS (Quantum MAterials Simulator)~\cite{S_qmas} with electronic wavefunctions expressed as two-component spinors~\cite{S_kosugi11} based on the projector augmented-wave method~\cite{S_blochl94,S_kresse99}.
The plane-wave energy cutoff is set to 40 Ry, and the $6\times 6\times 6$ $\bm{k}$-mesh is employed.
Te is wrongly described to be a metal in the LDA. 
In order to correct the band gap, we evaluate the $GW$ self-energy~\cite{S_hedin65}
using the full-potential linear muffin-tin orbital code~\cite{S_schilfgaarde06,S_miyake08}.
In the GW calculation, the spin-orbit interaction is neglected.
The $6\times 6\times 6$ $\bm{k}$-points are sampled and $51\times 2$ unoccupied conduction bands are included, where $\times 2$ is the spin degrees of freedom.
The calculation is sufficiently converged within 10 meV, and it does not affect our conclusions, except for slight deviation of transition pressures.
Fully-relativistic electronic structures including both the spin-orbit interaction (SOI) and $GW$ self-energy correction
are obtained by diagonalizing the following Hamiltonian: 
\begin{eqnarray}
%\begin{equation}
H^{\text{$GW$+SO}}_{mn}(\bm{R}) &=& \langle \phi _{m\bm{0}}|\mathcal{H}^{\text{LDA+SO}}|\phi _{n\bm{R}} \rangle 
\nonumber \\
&+& \langle \phi _{m\bm{0}}|-V_{\text{xc}}^{\text{LDA}}+\Sigma ^{GW} |\phi _{n\bm{R}} \rangle ,
\label{t}
%\end{equation}
\end{eqnarray}
where $\mathcal{H}^{\text{LDA+SO}}$, $V_{\text{xc}}^{\text{LDA}}$, and $\Sigma ^{GW}$ are 
the Kohn-Sham Hamiltonian in the LDA with the SOI, 
the LDA exchange-correlation potential, and the $GW$ self-energy, respectively.
$\phi _{n\bm{R}}$ is the maximally localized Wannier function (MLWF)~\cite{S_marzari97,S_souza01} 
of the $n$th orbital centered at the unit cell $\bm{R}$.
We construct 9$\times$2 MLWF's spanning the $p$ bands.
For each spin, three MLWF's out of 9 are centered at each of the three atoms in the unit cell. 
Experimental structures for the trigonal phase are used in the present study~\cite{S_keller77,S_adenis89,S_mccann72}.
The bond length for the intra- and inter-chain ($r, R$) is $(2.83, 3.49)$ \AA \ in Te 
and $(2.37, 3.44)$ \AA \ in Se, respectively.
Hence, Se has stronger one-dimensional character than Te.
The $a$ lattice constant decreases, whereas $c$ increases under pressure~\cite{S_keller77}.
At pressures between available experimental data in Ref.~\cite{S_keller77}, we determine the Hamiltonian in the MLWF basis by linear interpolation.

\noindent
\section*{S.2 Calculation of Monopole Charge}

To calculate the value of the monopole charge, we use a computational 
scheme proposed by Fukui \textit{et al.}~\cite{S_fukui05}.
We define a small cubic region around each degenerate point. 
A $U(1)$ link variable of the $n$th band is defined as
\begin{equation}
U_{\mu}(\bm{k}_l,s) = \langle n (\bm{k}_l,s)|n (\bm{k}_l+\hat{\bm{\mu}},s) \rangle / N_{\mu}(\bm{k}_l,s) , 
\label{umu}
\end{equation}
where $\mu =1, 2$ is the direction of the $\bm{k}$ vector on the surface of the cubic,
$\bm{k}_l$ is the $\bm{k}$ vector at the $l$th mesh point,
$s=1$-$6$ is the index for the six surfaces of the cubic,
$\hat{\bm{\mu}}$ is the vector between the nearest mesh points for the $\mu$ direction,
and $N_{\mu}(\bm{k}_l,s)= \langle n (\bm{k}_l,s)|n (\bm{k}_l+\hat{\bm{\mu}},s) \rangle $.
The monopole charge $\tilde{c}_n$ is calculated from $U_{\mu}(\bm{k}_l,s)$,
\begin{equation}
\tilde{c}_{n} = \frac{1}{2\pi i}\sum_{l,s}[\tilde{F}_{12}(\bm{k}_l,s)
-\Delta_1 \tilde{A}_2(\bm{k}_l,s)+\Delta_2 \tilde{A}_1(\bm{k}_l,s)] ,
\label{cn}
\end{equation}
where $\tilde{F}_{12}(\bm{k}_l,s)$ is a lattice field strength,
\begin{equation}
\tilde{F}_{12}(\bm{k}_l,s) = \ln U_1(\bm{k}_l,s)U_2(\bm{k}_l+\hat{\bm{1}},s)U_1(\bm{k}_l+\hat{\bm{2}},s)^{-1}U_2(\bm{k}_l,s)^{-1} ,
\label{f12}
\end{equation}
\begin{equation}
-\pi < \frac{1}{i}\tilde{F}_{12}(\bm{k}_l,s) \leq  \pi ,
\label{f12rng}
\end{equation}
and $\Delta _{\mu} \tilde{A}_{\nu}(\bm{k}_l,s)$ is a difference of the gauge potential,
\begin{equation}
\Delta _{\mu} \tilde{A}_{\nu}(\bm{k}_l,s) =\ln U_{\nu}(\bm{k}_l+\hat{\bm{\mu}},s)- \ln U_{\nu}(\bm{k}_l,s) ,
\label{amu}
\end{equation}
\begin{equation}
-\pi < \frac{1}{i} \tilde{A}_{\mu}(\bm{k}_l,s) \leq  \pi .
\label{amurng}
\end{equation}
We employ the $20 \times 20$ $\bm{k}$-mesh on the six surfaces of the cube,  
which we found is sufficient for numerical convergence. 

\noindent
\section*{S.3 Protection of Weyl Node and Symmetry Breaking}

The Weyl point between H and K in Te/Se belongs to the two-dimensional irreducible representation,
and in that sense it originates from symmetry.
At the same time it is protected topologically due to its nonvanishing monopole charge.
Therefore, even when the spatial symmetry is lowered, the three-dimensional Weyl point does not disappear, thanks to topological protection~\cite{S_Volovik,S_Murakami07b}.
The Weyl node only shifts in momentum space to another $\bm{k}$ point having lower symmetry.

\begin{figure}[ptb]
\centering 
\includegraphics[clip,width=0.45\textwidth ]{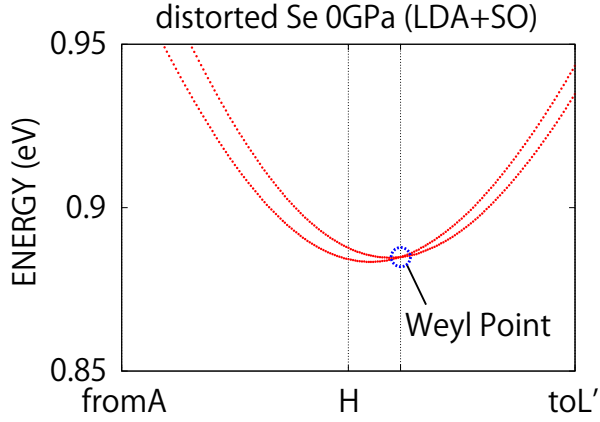}
\caption{
(Color online)
Electronic band structure of distorted Se having only $C_2$ symmetry along a axis.
The length between the B site and the center axis of the Se chain is 0.95 times smaller than that of the other sites.
The energy is measured from the Fermi level.
The calculation is based on the LDA+SO.
}
\label{bndc2}
\end{figure}

For example, if only the B site in Te/Se moves to the center axis of the Te/Se chain, the $S_3$ symmetry is broken and one of the $C_2$ rotations survives.
Then, the Weyl node of H$_6$ moves along the A-H-L' line having the $C_2$ symmetry (Fig.~\ref{bndc2}).
When the last $C_2$ symmetry is also broken,
the Weyl node will survive, and it will move to another $\bm{k}$ point with no additional symmetry.

\noindent
\section*{S.4 Topological Surface State}

We calculate the surface state on the $(01\bar{1}0)$ surface of Te in the Weyl semimetal phase.
The calculation is based on the surface Green's function method~\cite{S_turek97,S_dai08}, 
using the \textit{ab initio} parameters of Te at 2.18 GPa in the $GW$+SO method for this calculation.
The system contains 240,000 atoms in total.
The surface Green's function is the projection of the full Green's function to that of 12 atoms in the surface region.
$101\times 101$ $\bm{k}$-mesh and $101$ energy mesh are employed near the Weyl points.

\begin{figure*}[ptb]
\begin{center}
\includegraphics[clip,width=0.95\textwidth ]{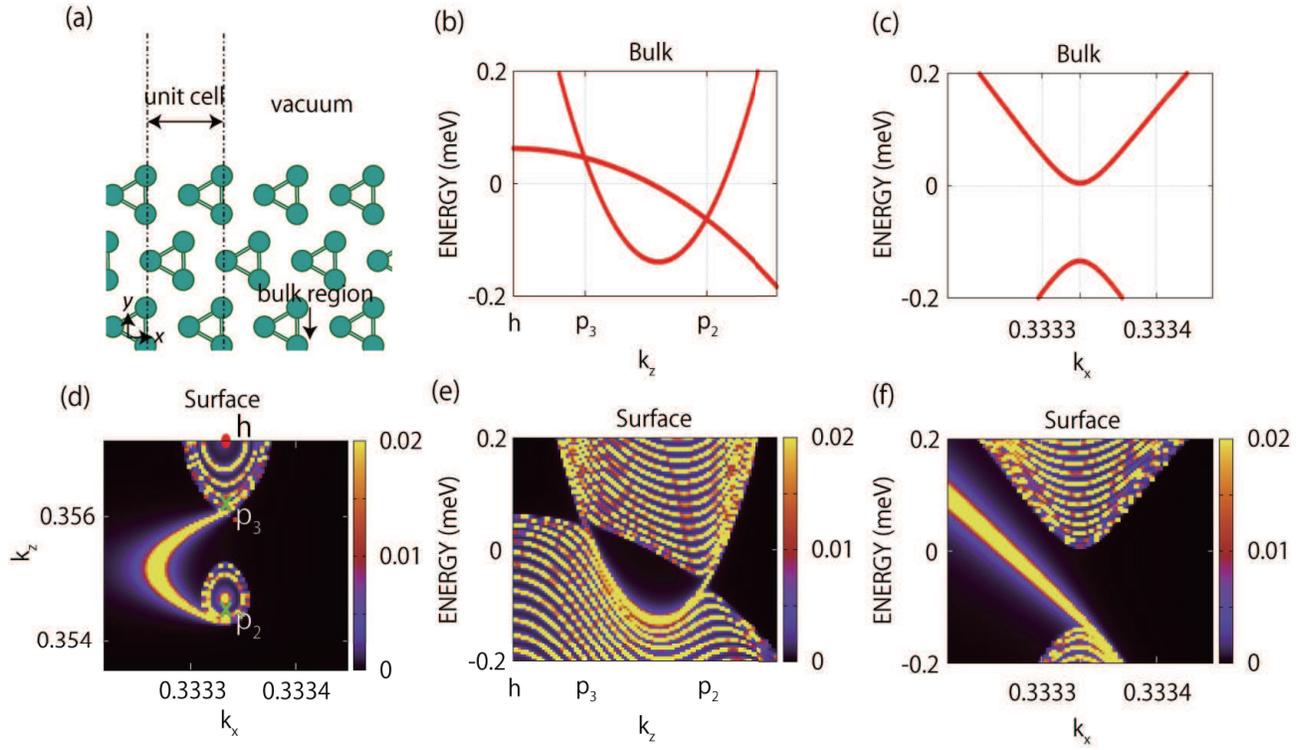}
\end{center}
\caption{
(Color online)
(a) Crystal structure of Te near $(01\bar{1}0)$ surface.
(b) Electronic band structure of bulk Te along the $k_z$ direction, where we take three unit vectors as $(a,0,0)$, $(0,\sqrt{3} a, 0)$, and $(0, 0, c)$.
h, p$_2$, and p$_3$ correspond to H, P$_2$, and P$_3$ in the normal unit cell.
(c) Electronic band structure of bulk Te along the $k_x$ direction, where $k_z$ is that of midpoint between p$_2$ and p$_3$.  
(d) The intensity color plot of the density of state at the Fermi level in the surface region.
(e) and (f) are the intensity color plot of the density of state in the surface region corresponding to (b) and (c), respectively.
}
\label{sgf}
\end{figure*}

Figure~\ref{sgf}(a) is the crystal structure near the $(01\bar{1}0)$ surface.
For comparison, we also show the correspondent band structures of the bulk in Figs.~\ref{sgf}(b)(c).
The Weyl points are at p$_2$, and p$_3$, and the gap is opened between these two Weyl points.

In the Weyl semimetal phase, the nontrivial surface states emerge on the surface as the consequence of the topological nature of the bulk. 
Figure~\ref{sgf}(d) shows the density of states at the Fermi energy on the surface.
The hole/electron pocket originating from the bulk states exists around the h/p$_2$ point.
The folded bulk states are on the inside of the band structure of the bulk (Figs.~\ref{sgf}(e)(f)).
In addition to the bulk states, the Fermi arc connecting the two Weyl points (p$_2$ and p$_3$) exists in $k_x \leq 1/3$ region.
The topological surface states run between the valence and conduction bands,
and those at the Fermi level are the Fermi arc in Fig.~\ref{sgf}(d).

% 
%\begin{figure}[ptb]
%\centering 
%\end{figure} 
%

\noindent
\section*{S.5 Spin texture around a Weyl node}

Here we show that spin texture around a Weyl node is nontrivial, having a
nonzero Pontryagin index. An example is a hedgehog spin texture discussed in 
the main text.
Suppose a Weyl node at $\bm{k}_0$. An effective Hamiltonian around the 
Weyl node for the two 
bands involved is represented as \cite{S_Volovik,S_Murakami07b}
\begin{equation}
H=\sum_{i,j=1,2,3}a_{ij}\delta k_{i}\sigma_{j}= \sum_j b_j \sigma_j,\ 
\ b_j=\sum_{i}a_{ij}\delta k_{i}
\label{eq:Heff}
\end{equation}
where $a_{ij}$ is a constant, 
$\delta \bm{k}=\bm{k}-\bm{k}_0$ is the deviation of the wavevector from 
the Weyl node at $\bm{k}_0$, and $\sigma_j$ is the Pauli matrix acting on 
a pseudospin space.
It is straightforward to 
calculate its eigenvectors $|\psi_{\pm}\rangle$ for the eigenvalues $E=\pm|\bm{b}|$, 
corresponding to the upper and lower bands. 
Then the monopole charge $N_{\pm}$ for the Weyl node for each band with $E=\pm|\bm{b}|$ is
calculated as a surface integral of the Berry curvature over a closed surface 
enclosing the Weyl node:
\begin{equation}
N_{\pm}=\iint \frac{dS_k}{2\pi}\  i \varepsilon_{ijk}
\left\langle \frac{\partial \psi_{\pm}}{\partial k_i}\right|\left.
\frac{\partial \psi_{\pm}}{\partial k_j}\right\rangle=\mp {\rm sgn}({\rm det} a),
\end{equation}
where ${\rm det}a$ is the determinant of the matrix with entities $a_{ij}$. We here assume ${\rm det} a\neq 0$, because otherwise it is not a Weyl node. 
On the other hand, 
the expectation value of the pseudospin for each band is 
\begin{equation}
\bm{S}_{\pm}\equiv\langle\psi_{\pm}|\bm{\sigma}|\psi_{\pm}\rangle
=\pm \bm{b}/|\bm{b}|,
\end{equation}
which depends on $\bm{k}$. Therefore, the Pontryagin index for the spin texture around the Weyl point 
is 
\begin{equation}
P_{\pm}=\iint \frac{dS_k}{4\pi|\bm{S}_{\pm}|^3} \ \varepsilon_{ijk}
\bm{S}_{\pm} \cdot 
\left(\frac{\partial \bm{S}_{\pm}}{\partial k_i}\times
\frac{\partial \bm{S}_{\pm}}{\partial k_j}
\right)=\pm {\rm sgn}({\rm det} a).
\end{equation}
Hence we conclude that $P_{\pm}=-N_{\pm}=\pm 1$. meaning that the 
Pontryagin index for the spin texture around the Weyl node is $\pm 1$ which is
equal to $(-1)$ times the monopole charge.

In general condensed matter systems, various bands with various orbitals are involved in the Weyl 
node. In this case the Pauli matrix in Eq.~(\ref{eq:Heff}) acts on a pseudospin space, not a spin itself. 
Nevertheless, in the present system with a spin-orbit coupling without inversion symmetry, every 
state is generally nondegenerate, and has a specific spin state. Hence the pseudospins are closely related with 
real spins, and therefore the spin texture for real spins is also expected to have nontrivial spin structure 
with a nontrivial Pontryagin index. As we found in the first-principle calculations, it is indeed the case for Te.

%%%%%%%%%%%%%%%%%%%%%%%%%%%%%%%%%%%%%%%%%%%%%%%%%%%%%%%%%%%%%%%%%%%%%%%%%%%%%%


\begin{thebibliography}{99}
\bibitem{se14gpa} In the calculation of Se at 14.0 GPa, however, we exceptionally use the internal atomic coordinate $u$ at ambient pressure according to Ref.~\cite{starkloff78}, because there are no experimental data for Se under pressure.
\end{thebibliography}
\end{document}